\documentclass[letter]{aa}

\usepackage{txfonts}
\usepackage{graphicx}

\begin{document}

  \title{Unexpected radial trend of the iron abundance in a sample of monometallic Galactic globular clusters}

      \author{V. Kravtsov\inst{1,2}}

\offprints{V. Kravtsov}

   \institute{Instituto de Astronom\'ia, Universidad Cat\'olica del Norte,
              Avenida Angamos 0610, Antofagasta, Chile\\
              \email{vkravtsov@ucn.cl}
            \and
              Sternberg Astronomical Institute, Lomonosov Moscow State University, University Avenue 13,
              119992 Moscow, Russia\\
             }

   \date{Received xxxxx / Accepted xxxxx}

   \abstract
{} {We study the relationship between the iron abundance (IA) in red giant branch (RGB) stars and their radial
distribution (RD) in Galactic globular clusters (GCs).} {We relied on publicly available archival data on
IA in red giants (RGs) of GCs. We built a sample of ten target GCs in which the number of these RGs exceeded
one hundred stars. They span a wide range of projected radial distance (PRAD) in their parent GCs.}
{In each GC of the sample, we compared the RDs of two sub-samples of stars, more iron-rich (IR) and more
iron-poor (IP) than the clusters' mean values of [Fe/H]. Their RDs turned out to be different at
statistically significant confidence levels in NGC 104 (47 Tuc), NGC 1851, NGC 3201, and NGC 6752 in the
sense that the IP RGs were more centrally concentrated than their IR counterparts. In 47 Tuc, the
difference is significant at a higher confidence level within the PRAD of $R \approx 8\farcm0$, where
the IA increases by $\Delta$[Fe/H]$ \sim$ 0.03 dex toward the cluster outskirts. In the latter three GCs,
$\Delta$[Fe/H]$ \sim$ 0.05 dex. Interestingly, the $V$ magnitude of the RGB bump and the horizontal branch
was recently shown to fade outward in 47 Tuc and was suggested to originate from a He abundance trend. We
estimated the fading caused by the IA trend. It is similar to that observed for the RGB bump.
Although the difference between the RDs of IP and IR RGs is statistically insignificant in other GCs,
NGC 288 is the only GC of the sample, in which IR RGB stars are formally more centrally concentrated. 
We checked whether the trend could be caused by a possible spurious effect, in particular due to
systematically brighter IP than IR RGs. NGC 3201 was the only GC where difference between the RDs of
IP and IR RGs became insignificant after corrections were applied. The latest data on the IA in a
sample of RGs in NGC 3201 confirmed that IP RGs are clearly more centrally concentrated. However,
a spurious nature of the trend cannot be fully ruled out. Our results imply that the unusual radial trend
of IA in GCs, if real, may occur fairly frequently in GCs with an internally small scatter of the [Fe/H] ratio. 
Interestingly, three of the four GCs are highly concentrated.} {}

   \keywords {globular clusters: general --
                globular clusters: individual: NGC 104, NGC 288, NGC 1851, NGC 3201, NGC 4590, NGC 5904, NGC 6121, NGC 6254, NGC 6809, NGC 6752}

\maketitle

\section{Introduction}
\label{introduc}

There is a growing body of evidence that there are multiple stellar populations
(MSPs) in a growing number of Galactic globular clusters (GCs). The evidence comes either from
the manifestation of different photometric effects in color-magnitude diagrams (CMDs),
incompatible with simple populations, or from spectroscopically estimated elemental
abundance variations among cluster stars that are not caused by evolutionary effects in
the stars. A separate overview of the dedicated publications is beyond the scope of this
Letter. Only some relevant results and papers are referred to elsewhere.
For more details we refer to the overview provided in Gratton et al. (\cite{gratetal12}).

This study was mainly motivated by recent findings of radial segregation between sub-populations
of red giant and sub-giant branch (RGB, SBG) stars, which differ in their photometric characteristics,
primarily in UV-based ones, in several GCs. In particular, Kravtsov et al.
(\cite{kravtsovetal10a,kravtsovetal10b}) and Kravtsov et al. (\cite{kravtsovetal11}) reported on
this segregation in NGC 3201 (see also Carretta et al. \cite{carrettal10b}),
NGC 1261, and NGC 6752, respectively. Similar results were obtained
for RGB stars in a number of GCs by Lardo et al. (\cite{lardoetal11}), relying on multi-color SDSS
photometry including in the UV-band, in which the radial effect was detected, but not in other
bands. Moreover, in addition to the radial trends observed in ultraviolet photometric bands, Nataf et
al. (\cite{natafetal11}) found statistically significant radial variations of the brightness in the
RGB bump position and horizontal branch (HB) level in GC NGC 104 (47 Tuc) and argued that these variations
could be caused by a radial trend of the He abundance in the cluster. Milone et al. (\cite{milonetal12})
found evidence of two populations and a radially changing proportion between them in the same GC.
Accordingly, if the radial photometric trends are not the result of some spurious effects, they are
probably caused, in general, by a superposition of various factors contributing to the revealed
variation of brightness in the UV part of the spectrum. Among the most probable contributors
can be molecules that contain some key chemical elements (see details in Sbordone et al. \cite{sbordonetal11}).
Given the well-known (anti)correlations between the key elements in stars of GCs, there is need to study
the radial dependence of their abundances in the GCs.

\begin{table*}
\caption{Data for the total samples and sub-samples of RGs in the selected GCs}
\label{gcparam}
\centering
\begin{tabular}{cccccccccc}
\hline \hline \noalign{\smallskip}
Cluster  & \multicolumn{4}{c}{Total sample of RGs} & \multicolumn{2}{c}{IR sub-sample} & \multicolumn{2}{c}{IP sub-sample} & KS test \\

NGC  & [Fe/H]$_{I}$ & $\sigma$ & N$_{tot}$ & $R_{out}(\arcsec)$ & N$_{IR}$ & $\overline{R}_{IR}(\arcsec)$ &  N$_{IP}$ & $\overline{R}_{IP}(\arcsec)$ & P(\%) \\
\hline \noalign{\smallskip}
104  & -0.743 & 0.032 & 147 & 734.5 & 78 & 294.6 & 69 & 247.9 & 95.1 \\
288  & -1.219 & 0.042 & 110 & 739.9 & 56 & 172.0 & 54 & 215.0 & 73.8 \\
1851 & -1.158 & 0.051 & 124 & 636.3 & 61 & 224.2 & 63 & 167.3 & 99.6 \\
3201 & -1.495 & 0.049 & 149 & 554.0 & 72 & 262.8 & 77 & 183.7 & 99.8 \\
4590 & -2.227 & 0.071 & 122 & 610.5 & 60 & 206.5 & 62 & 178.6 & 72.8 \\
5904 & -1.346 & 0.023 & 136 & 694.8 & 70 & 257.1 & 66 & 229.3 & 76.9 \\
6121 & -1.200 & 0.025 & 103 & 693.4 & 49 & 223.9 & 54 & 223.6 & 9.5 \\
6254 & -1.556 & 0.053 & 147 & 611.4 & 72 & 208.5 & 75 & 189.8 & 40.9 \\
6752 & -1.562 & 0.041 & 137 & 620.6 & 71 & 266.3 & 66 & 187.0 & 99.4 \\
6809 & -1.967 & 0.044 & 156 & 595.7 & 75 & 230.9 & 81 & 198.4 & 92.0 \\
 \hline
\end{tabular}
\end{table*}

\begin{figure}
\centerline{
\includegraphics[clip=,angle=-90,width=9.0 cm,clip=]{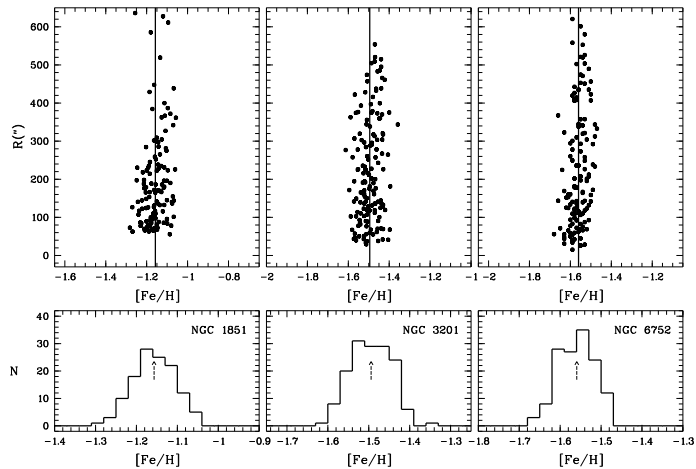}}
\caption{Upper panels show [Fe/H]$_{I}$ - PRAD diagrams and lower panels are metallicity distribution histograms in NGC 1851, NGC 3201, and NGC 6752. The continuous lines and arrows mark the mean values of [Fe/H]$_{I}$ in the upper and lower panels, respectively.}
\label{radtrend}
\end{figure}

The main goal of this paper is to study the radial distribution (RD) of RGB stars distinguished by their
iron abundance (IA) in GCs. We focus on the GCs that were characterized as "monometallic" in Carretta et al.(\cite{carrettal09}) because of the typically low dispersion of the [Fe/H] ratio they found
from massive measurements of IA for large samples of RGB stars in the GCs.

\section{Data and calculated quantities}

We used the possibilities provided by the Virtual Observatory, namely the VizieR
catalog access tool, which relies on publicly available data of high-resolution spectroscopy
of large samples of RGB stars in a diversity of GCs. The spectroscopy was made using FLAMES-GIRAFFE
spectra and respective results were published in Carretta et al. (\cite{carrettal07,carrettal09,carrettal11}).
We refer to these key papers for the details about observations, data reductions, and other
points concerning the data, including the main conclusions reached from the analysis of the data. We selected
only those GCs from their list that contained more than one hundred RGB stars
(i.e., N$_{tot} > 100$) with available measurements of the abundance of neutral iron, [Fe/H]$_{I}$. We
finally were able to build a sample of ten GCs. Using the values of IA of individual stars
in each GC, we calculated the mean values of [Fe/H]$_{I}$ and standard deviation (rms), $\sigma$, using relevant commands of MIDAS system. The total sample of RGB stars, N$_{tot}$, in each GC was divided by sub-samples of more iron-poor (IP, N$_{IP}$) and more iron-rich (IR, N$_{IR}$) RGB stars with IA lower and higher than its mean value in each GC, respectively. Figure~\ref{radtrend} demonstrates and explains that and other details, with specific
reference to three of four GCs of particular attitude explained below. It shows the RD of IA with respect to the mean value of [Fe/H]$_{I}$ and metallicity distributions histograms for the samples of RGB stars in these GCs. For the sub-samples of IP and IR RGB stars we calculated their mean projected radial distances
(PRAD; expressed in arcseconds) in the parent GCs, $\overline{R}_{IP}$ and $\overline{R}_{IR}$, respectively.
Finally, the probability (expressed in percent) P that the two sub-samples have different RDs in their parent
GCs was estimated by applying a Kolmogorov-Smirnov (K-S) test. All these data on the selected GCs and their
sub-samples of IP and IR RGs are listed in Table~\ref{gcparam}.

\begin{figure*}
\centerline{
\includegraphics[clip=,angle=-90,width=5.2 cm,clip=]{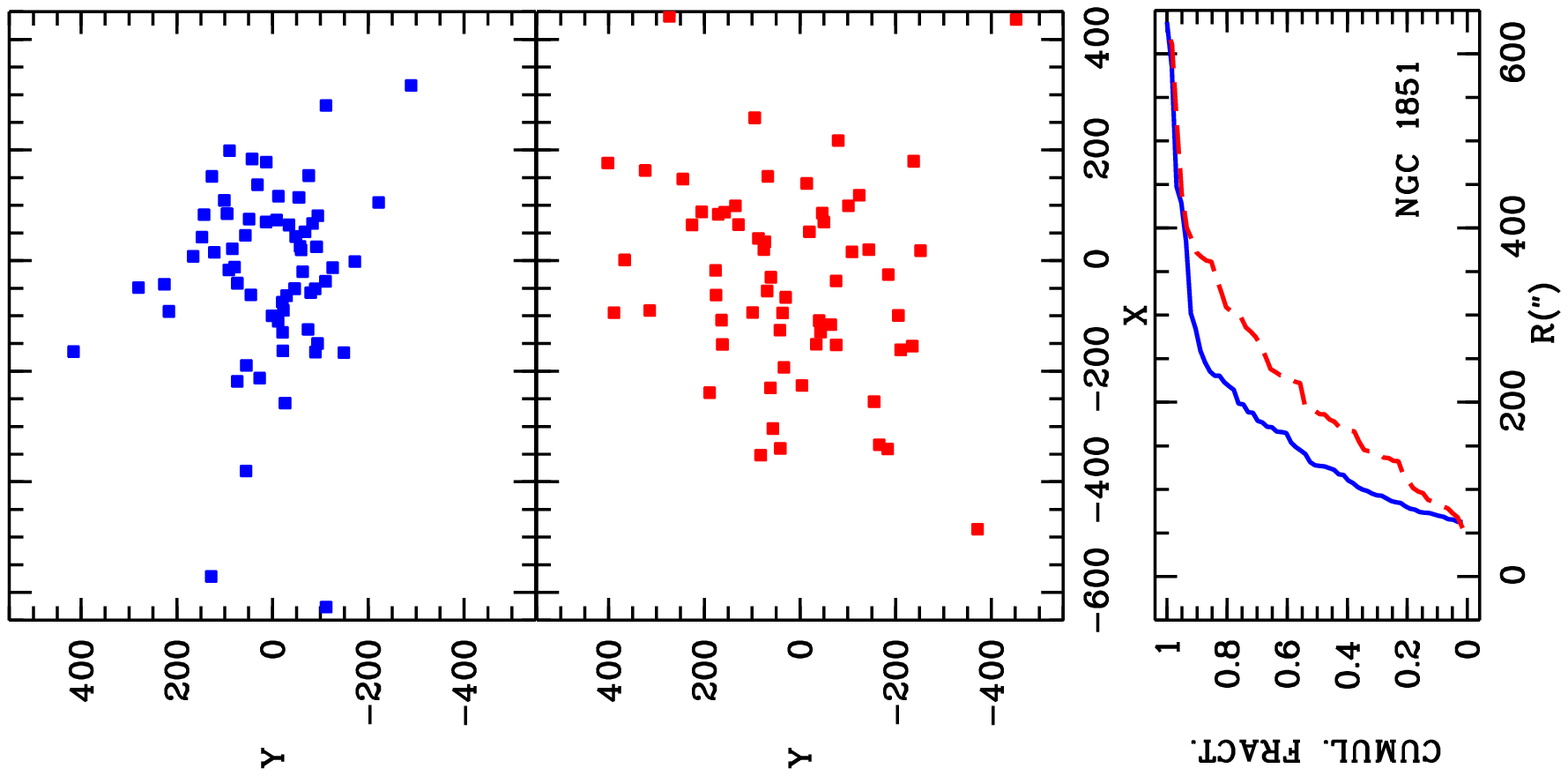}
\includegraphics[clip=,angle=-90,width=4.2 cm,clip=]{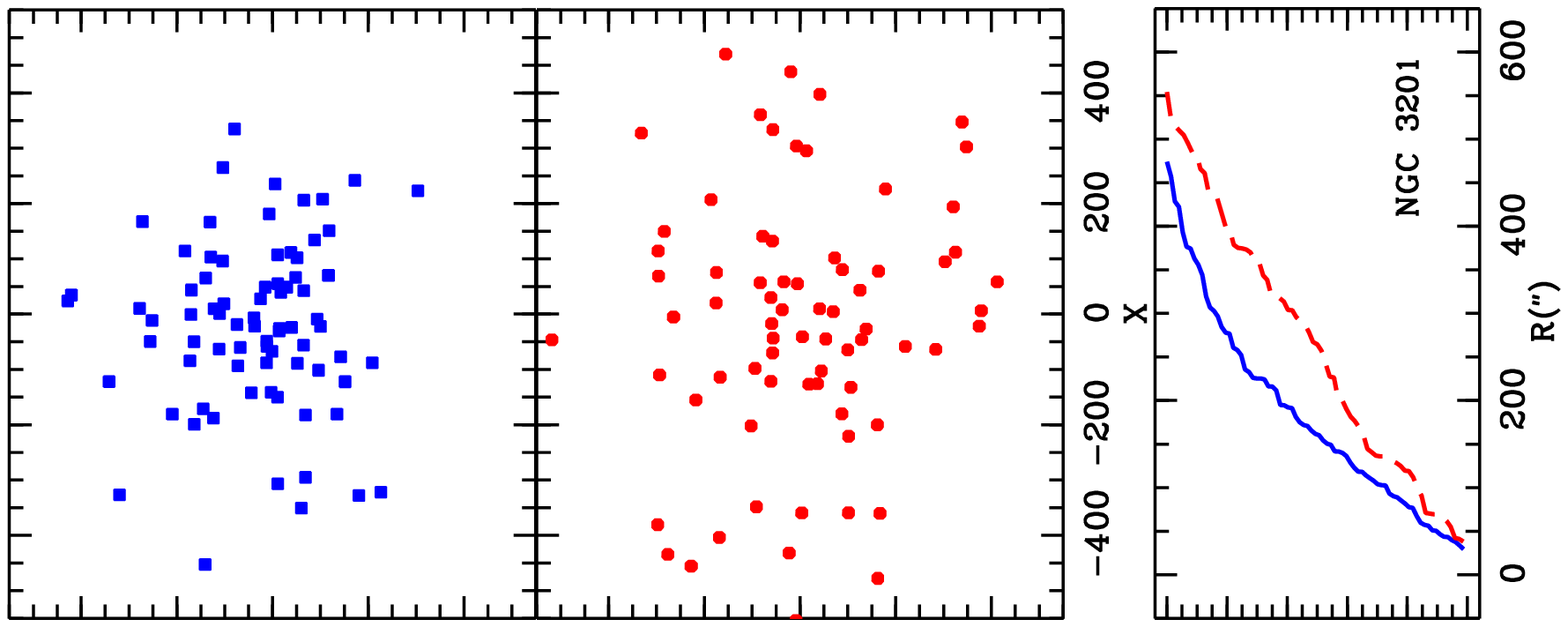}
\includegraphics[clip=,angle=-90,width=4.2 cm,clip=]{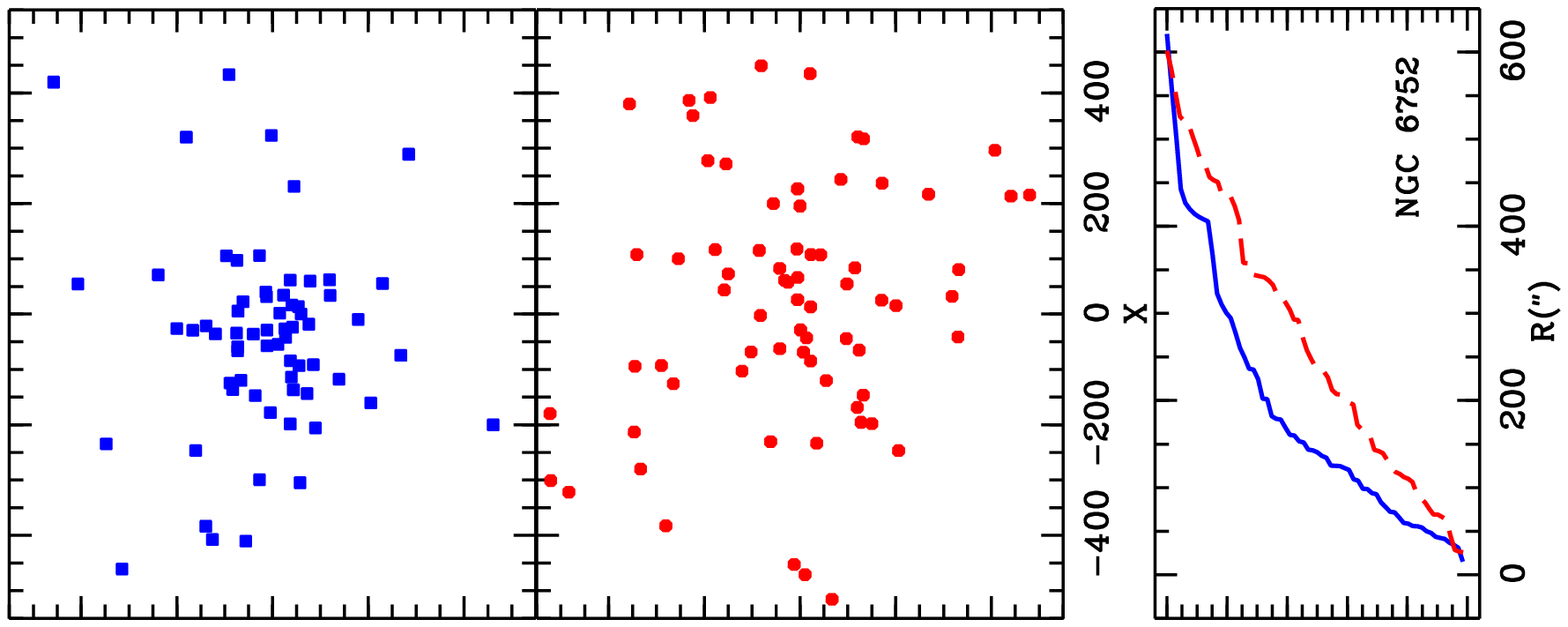}}
\caption{Upper and middle panels show the position of IP (blue dotes) and IR (red dotes) RGB stars in the fields of the three GCs. Lower panels are cumulative RDs for IP (blue continuous line) and IR (red dashed line) RGs. PRAD from the cluster centers, ($R$), and the rectangular coordinates X and Y (whose axes are aligned along the $\alpha$ and $\delta$ axes, respectively) with respect to the centers of the clusters are expressed in arcseconds.}
\label{posinfield}
\end{figure*}

\section{Results and final comments}
\label{sgb}

\noindent
$\bullet$ First, the K-S tests gave very high probabilities, P $>$ 99\%, that IP and IR RGs have essentially different RDs in three GCs, NGC 1851, NGC 3201, and NGC 6752. The location of the RGs of each sub-sample in the cluster fields and corresponding cumulative RDs are shown in Figure~\ref{posinfield}. They demonstrate that the difference of the RDs are the following: IP giants are more centrally concentrated than their IR counterparts. In other words, there is a statistically significant radial trend of IA in these GCs, expressed in the lower [Fe/H]$_{I}$ ratio in the central parts of the GCs. 47 Tuc is the fourth GC of our sample where the radial trend of IA in this sense is very probable: it is marginally statistically significant, i.e., at P = 95.1\% confidence level. The probability of more centrally concentrated IP RGB stars is fairly high (P = 92\%) in NGC 6809 (M55). However, strictly speaking, it is statistically insignificant. The lowest probability of any difference between RDs of IP and IR RGB stars is in NGC 6121 (M4), P = 9.5\%. This agrees with P = 28.5\%, another estimate we made using the data of Marino et al. (\cite{marinoetal08}) from FLAMES/UVES spectroscopy of a sample of 105 RGs in the same GC. In the remaining GCs of the sample, the probability of such a difference varies at intermediate level, implying that the difference is statistically insignificant, too. Despite this, we note that NGC 288 is formally the only GC of the sample, in which IR RGB stars are more centrally concentrated.

\noindent
$\bullet$ Second, we estimated the radial change of the IA in the four GCs. The difference between the central, lowest, value of the IA and its highest value achieved in the outer region of each cluster was found to be around $\Delta$[Fe/H] $\approx 0.050 \pm0.013$ dex in NGC 1851, NGC 3201, and NGC 6752. In 47 Tuc, it is a factor of $\sim1.5$ lower: $\Delta$[Fe/H] $\approx 0.030 \pm0.013 $ dex. The lower, central value of the IA in 47 Tuc is somewhere within PRAD of $R \approx 70\arcsec$, and it gradually increases at larger PRAD. The strongest difference in IA with respect to the central value of the [Fe/H]$_{I}$ ratio is achieved somewhere at PRAD $R \approx 450\arcsec$. The difference between RDs of the sub-samples of IP and IR RGB stars in 47 Tuc is statistically significant at a higher confidence level within the same PRAD, $R \approx 480\arcsec$, than in the total range of PRAD spanned by the RGB stars of the sample. We remark that the estimated magnitude of the radial trend of the IA in the four GCs is fairly low, of the same order as the rms errors quoted for the measurements of the [Fe/H] ratios in these GCs.

\noindent
$\bullet$ Third, given that the radial trend of the IA is weak and fairly unusual in terms of the unexpected decreasing IA toward the centers of GCs, we suspect that it might be caused by some kind of spurious systematic effect. In particular, we found that IP RGs are systematically brighter (therefore cooler) than their IR counterparts. This trend could be caused by a systematic error (if any) in estimating [Fe/H] at different temperature. The difference in mean magnitudes between the two sub-populations is typically of a few tenths of magnitude. We examined the luminosity functions (LFs) of the sub-populations of IP and IR RGs and imposed a constraint onto the original samples of RGs. We excluded not only the brighter parts of the RGB LFs, but also their fainter ends. The imposed upper and lower limits, $V_{sup}$ and $V_{inf}$ are listed in Table~\ref{RGaftercorrect} along with other parameters (the same as in Table~\ref{gcparam}) calculated for new (sub-)samples of RGs in the three GCs of principal interest, 47 Tuc, NGC 3201, and NGC 6752 and additionally in NGC 6809. New mean values of [Fe/H]$_{I}$ were redetermined for the reduced samples of RGs and then used to redefine the sub-samples of IP and IR RGs. No constraint was imposed on the original sample of RGs of NGC 1851, because the difference between the mean $V$ magnitudes of its IP and IR RGs, $\Delta V < 0.1$ mag, was the least compared with the original samples of other three GCs. In 47 Tuc and NGC 3201, the residual difference remained after the corrections. In NGC 6752, the difference was reduced and became of same order of magnitude, i.e. $\Delta V < 0.1$ mag, but in the opposite sense: IR RGs became, on average, slightly brighter than IP ones. As seen in Table~\ref{RGaftercorrect}, despite the significantly reduced sample size of RGs, the difference between $\overline{R}_{IR}$ and $\overline{R}_{IP}$ became even larger in 47 Tuc and NGC 6752 and remained statistically significant at high confidence level. In 47 Tuc, the confidence level became higher. We recall that NGC 3201 is a somewhat special case, since the reddening across the cluster face is widely known to be highly irregular and patchy. This means that, the constraint we applied, especially at the faint end of the LF, could lead to an artificial rejection of the most reddened RGs from the regions with a systematically larger proportion of IR RGs. It is worth noting that Simmerer et al. (\cite{simmereretal13}) have just published new data on the IA in a sample of 26 RGs of the GC. They used UVES and MIKE spectra of the target RGs (21 and 5 stars, respectively), which are of higher resolution than GIRAFFE spectra. The difference between the RDs of IP and IR stars is obvious. Simmerer et al. (\cite{simmereretal13}) have noted that the most IP stars are more centrally concentrated. According to our estimate, the difference between the RD of IP and IR RGs is statistically significant at high confidence level, P = 96\%, despite the fairly limited sample size. Still, a spurious nature of the revealed trend, which might be caused by some systematic effect, cannot yet be ruled out.

We also studied the large sample of RGs in NGC 6809. We found a notable difference between the LFs (in their brighter parts) of the original sub-samples of RGB stars and we rejected 55 RGs brighter than $V_{sup} = 13.60$ in the range of more that 2 mag. The redefined sub-samples of IP and IR RGs (with essentially reduced sample sizes) show even slightly larger differences between $\overline{R}_{IP}$ and $\overline{R}_{IR}$ than do the original sub-samples. The difference between their RD is significant at a confidence level of P = 82\%. The difference between the mean $V$ magnitudes of the two sub-samples decreased to $\Delta V = 0.09$  in the sense that the IP RGs remain, on average, brighter. Notice that the systematically brighter mean magnitude of IP RGs can in fact be interpreted as indirect evidence of the real difference in IA among RGs of the GCs under study. Indeed, when a sample of RGs is taken at random in a range of brightness, the LF of its IP sub-sample (including the RGB bump) is expected to be shifted brightward with respect to the LF of IR RGs.

\noindent
$\bullet$ Fourth, the radially fading of the red HB and RGB bump found by Nataf et al. (\cite{natafetal11}) and the variation of IA in 47 Tuc occur virtually in the same range of PRAD. Nataf et al. (\cite{natafetal11}) argued that the fading was caused by the radial variation of the He abundance quantified from its theoretically predicted effect on the HB and RGB bump brightness variations compatible with the observed ones. At the same time, Nataf et al. (\cite{natafetal11}) found evidence for a slight color variation of RGs that was not predicted by the models with He variation, which implies that "there may be an additional factor at play". The authors concluded that "the color variation would be consistent with the stars nearer the center either having a {\it lower} metallicity, $\delta$[Fe/H]$ \approx$ 0.05 dex, or a temperature colder by $\delta$T $\approx$ 17 K". This assumption perfectly agrees with our finding and evaluation of the radial trend of the IA. We estimated the radial variations both of the red HB level and RGB bump position in the $V$ magnitude, which might be caused by the radially increasing IA by $\Delta$[Fe/H] $\approx 0.030$ dex. With this aim, we applied the dependencies of M$_{V,HB}$ and M$_{V,RB}$ on [Fe/H] taken from Chaboyer et al. (\cite{chaboyeretal96}) [M$_{V,HB} = $0.2*[Fe/H] + 0.98] and Alves \& Sarajedini (\cite{alvsarad99}) [M$_{V,RB} =$ 0.85*[Fe/H] + 1.63], respectively. For our calculations we accepted the IA in the central and outer parts of 47 Tuc to be [Fe/H] = -0.76 and [Fe/H] = -0.73, respectively. We derived that the HB and RGB bump brightness would fade toward the outskirts of 47 Tuc by $\Delta V_{HB} =$ 0.006 mag and $\Delta V_{RB} =$ 0.026 mag. Therefore, the former value is negligible compared with the value found by Nataf et al. (\cite{natafetal11}), whereas the latter is similar to that deduced by these authors. Accordingly, the probable radial variation of the IA, evidence of which we find in 47 Tuc, may be at least one of the main contributors to the RGB bump fading, but it is presumably a negligible contributor to the HB brightness variation. This implies that the contribution of the assumed He abundance variation to the fading of the RGB bump is lower.

\begin{table}
\caption{Data of the total samples and sub-samples of RGs in four GCs after the corrections applied to the original samples}
\label{RGaftercorrect}
\centering
\begin{tabular}{ccccc}
\hline \hline \noalign{\smallskip}
Clus.  & \multicolumn{1}{c}{Total sample RGs} & \multicolumn{1}{c}{IR RGs} & \multicolumn{1}{c}{IP RGs} & KS test \\

NGC  & [Fe/H]$_{I}$ \ $V_{sup}$ \ $V_{inf}$ & N$_{IR}$ \ $\overline{R}_{IR}(\arcsec)$ &  N$_{IP}$ \ $\overline{R}_{IP}(\arcsec)$ & P(\%) \\
\hline \noalign{\smallskip}
104  & -0.740 \ 13.2 \ 14.6 & 60 \ \ 306.2 & 57 \ \ 248.0 & 99.5 \\
3201 & -1.500 \ 13.3 \ 16.2 & 60 \ \ 216.6 & 64 \ \ 182.9 & 60.0 \\
6752 & -1.554 \ 13.1 \ 14.6 & 58 \ \ 269.1 & 44 \ \ 179.4 & 98.9 \\
6809 & -1.958 \ 13.6 \ 15.4 & 53 \ \ 244.6 & 48 \ \ 207.2 & 82.0 \\
 \hline
\end{tabular}
\end{table}

\begin{figure}
\centerline{
\includegraphics[clip=,angle=-90,width=4.9 cm,clip=]{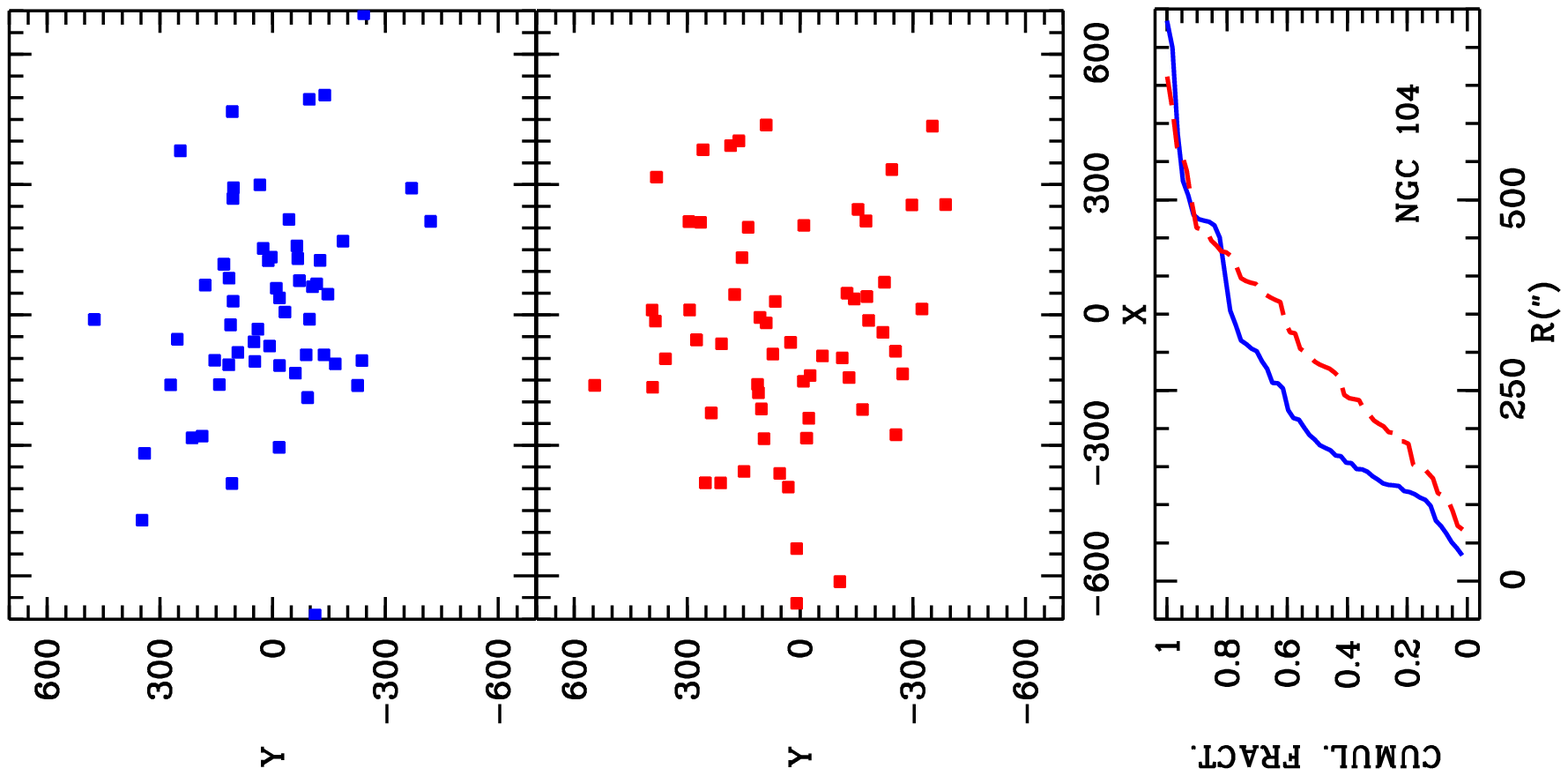}
\includegraphics[clip=,angle=-90,width=3.94 cm,clip=]{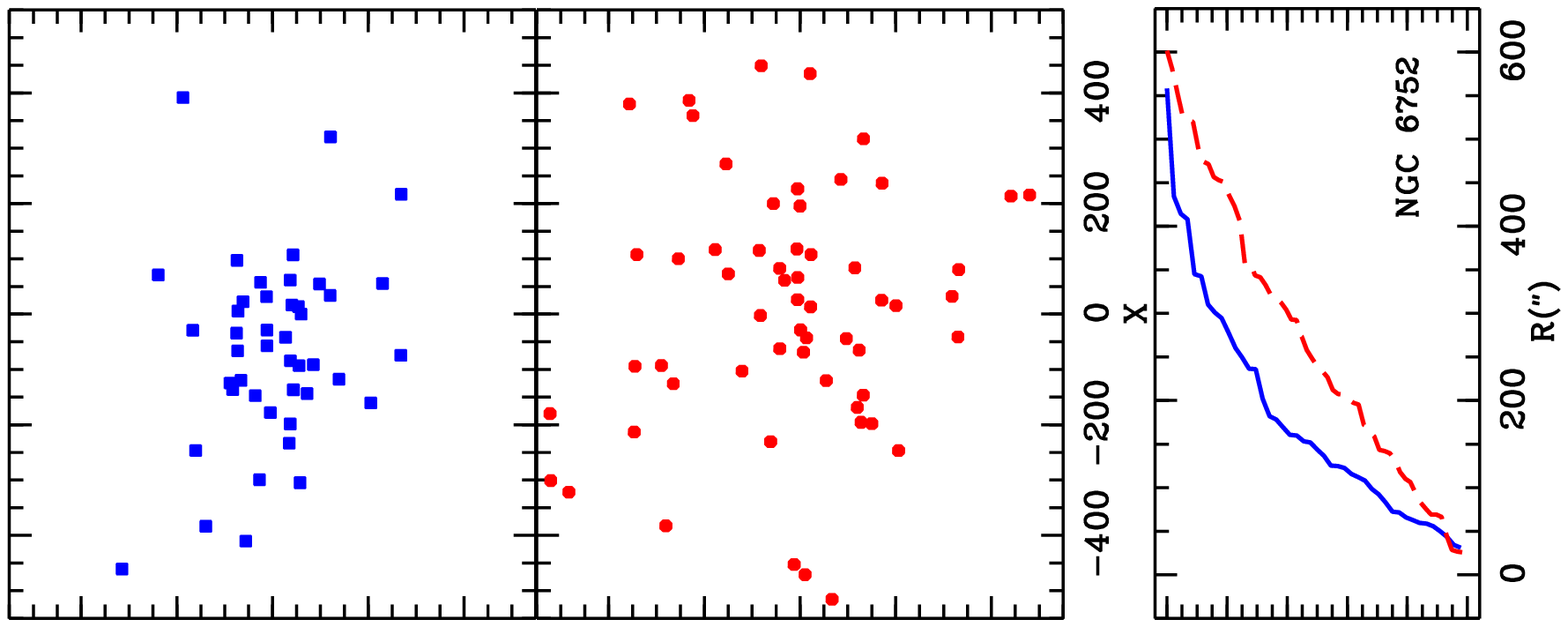}}
\caption{Same as in Figure~\ref{posinfield}, but for 47 Tuc and NGC 6752 after constraints imposed on the original data as described in the text.}
\label{posinfieldcorrect}
\end{figure}

\noindent
$\bullet$ Fifth, the radial segregation between the two sub-populations of RGs differing by IA in NGC 1851 was found previously by Carretta et al. (\cite{carrettal10a}). We note that our results agree well with theirs. Carretta et al. (\cite{carrettal10a}) argued that, among different alternatives, the merging of two GCs could better explain (reconcile) not only the variety of characteristics concerning the sub-populations of RGB and SGB stars, but also the morphology of the cluster HB. The results of our study imply that the decreasing IA toward the cluster center occurs relatively frequently among GCs. Interestingly, three of the four GCs with the most probable radial trend of IA, 47 Tuc, NGC 1851, and NGC 6752, are highly concentrated. In contrast, NGC 288 has a lower concentration and mass among the GCs of the sample. It is expected to be a less crowded GC (and M4, too).

\begin{acknowledgements}

This research has made use of the VizieR catalog access tool, CDS, Strasbourg, France. The author thanks the anonymous referee for useful comments that have improved the manuscript.

\end{acknowledgements}

\end{document}